 \relax
\documentstyle[12pt]{article}
\begin{document}
\renewcommand{\theequation}{\arabic{section}.\arabic{equation}}
\newcommand{\eqreset}{\setcounter{equation}{0}}

\vspace*{.9 in}
\begin{center}
{\large\bf PERSISTANT CURRENT IN ISOLATED MESOSCOPIC RINGS
   }

\vspace{.8 in}

{\sc J.F. Weisz$^{**2}$ ,R. Kishore $^2$}

 and

{\sc F.V. Kusmartsev$^{*1,3}$}

 \vspace{.5 in}

{\it Department of Theoretical Physics}
{\it  University of Oulu}
{\it SF-90570 ~ Oulu, ~ Finland$^1$}

\vspace{.5 in}

{\it Instituto Nacional de Pesquisas Espaciaes-INPE}
{\it C. P. 515, 12201-970, S. J.  Campos, Sao Paulo, Brasil. $^2$}

\vspace{.5 in}

    {\it Institute for Solid State Physics,University of Tokyo}
        {\it Roppongi,Minato-ku,Tokyo 106 $^3$}

(received ~~~~~~~~~~~~~~~~~~
\end{center}

\vfill
\eject
\begin{abstract}

Persistant current in isolated mesoscopic rings is studied using the
continium and tight-binding models of independent electrons. The
calculation is performed with disorder and also at finite temperature.
In the absence of disorder and at zero temperature agreement is
obtained with earlier results by D. Loss et. al., in that there is half
quantum flux periodicity for a large and odd number of electrons, but full
quantum periodicity for any even number of electrons in the
ring. Strong, disorder converts the period into full quantum periodicity.
Finite temperature  reduces the magnitude of the current,
 but preserves the quantum flux periodicity at zero
temperature. However the sign of
the current may change as disorder or temperature is increased.
A generalization of the parity effect,
previously  discussed by  Legett, Loss and Kusmartsev
is described for the case where there are electrons with spin,
influenced by finite temperature and disorder.

\end{abstract}

\vfill
\eject

\section{INTRODUCTION}

Since the early works by Byers  and Yang \cite{1} Kohn \cite{2} Bloch\cite{3}
 Gunther and Imry\cite{4}
Kulik \cite{5}
 Buttiker et al.\cite{6} \cite{7} and Cheung et al.\cite{8}-\cite{12}
it has been realized that spinless electrons do not lead to the same current
as electrons with spin, due to Fermi statistics. In reality the results
that were originally obtained, presumably for any number of electrons,
apply only to the cases of even number of electrons; or to the case of a
single electron. It was realized that there were four separate cases to
be analized, in which the number, $N$, of electrons on the ring can
take the values $N = 4n,\ 4n + 1,\ 4n + 2$ and $4n + 3$, where
$n = 0,1,2,3,\dots $.
Early realization of this situation is due to F.V. Kusmartsev\cite{13}
and calculations of current are found in D. Loss et al\cite{14}.
Based on this new information, a
more modern treatment of current in rings requires
adding disorder or using finite
temperature. In particular it was realized by D. Loss et al.\cite{14} that in
an ordered ring with a large enough odd number of electrons, persistent
current shows almost perfect half-quantum-flux periodicity rather than full
quantum flux periodicity of the spinless electrons. This result follows
simply from Fermi statistics for electrons with spin.

The behaviour of rings can be studied either for isolated rings or for
rings with leads. In the first case the number of electrons must  be a
constant, while in the second case electrons may be lost or gained through
leads. In the first cases the chemical potential is a function of flux,
calculated with $\sum_{n\alpha} f(E_{n_\alpha}) = N$ where the sum is
over energy levels and spins. Since the energy levels depend on flux, so does
chemical potential. In the case of ring with leads the chemical potential is
imposed from the outside, and must be considered a constant. Electrons are
then lost or gained through the leads as   with the increase of the flux
 the appropriate levels cross the
fixed chemical potential. Therefore the kind of averaging that  is required
depends on the physical condition of the rings, wheather they are isolated or
not. This work deals with isolated rings, without inelastic scattering.

The investigation of the influence of the spin degrees of freedom on the
Aharanov-Bohm effect is also especially important if spin-orbit interaction
is taken into account\cite{15}. This creates extra quasi-half flux periodicity
effects. It is well known that such interaction plays a major roll in
semiconductors, when electrons or holes interact with magnetic field.
Instead of paramagnetic resonance, we have the so-called combined resonance
(Rashba effect). The  orbital moments ot orbital currents on the ring,
   which
change in the magnetic  field, may also cause spin flip processes\cite{16}
via the spin-orbit interaction. However, we do not carry out such spin-orbit
calculations here, and neglect Zeeman splitting, which is a very small
effect for the fields of interest.
\bigskip

\section{METHODS OF CALCULATION}

We consider a system of $N$ non interacting electrons of mass $m$ on a
one dimensional ring of length $L$, threaded by a magnetic flux
$\phi$. The Hamiltonian of the system is given as

\begin{equation}
H = {1\over {2m}} \left [-i\hbar { tial \over { tial x}} +
e A\right ]^2 \rightarrow {\hbar^2\over {2m}} \sum^N_{\scriptstyle\alpha=1
\atop\scriptstyle\sigma =\pm} (-i { tial\over { tial x_\alpha}} -
{{2 \pi} \over L} f)^2
\end{equation}

\noindent where $x_{\alpha}$ is coordinate of $\alpha^{th}$ electron
of spin $\sigma$, along the ring, starting from some origin,
$f = \Phi /\Phi_0$  and $\Phi_0$ is the quantum flux ${h \over e}$ and
$\Phi$ is the total magnetic flux through the ring.
Using the wave function

\begin{equation}
\psi (x) \ = \ A e^{i k x} e^{ i {x \over L} 2 \pi f}
\nonumber
\end{equation}

\noindent and the boundary condition coming from the uniquiness of the
wave function, $\psi (x) = \psi (x + L)$, one obtains the single
particle energy levels.

\begin{equation}
E_{n \sigma} = {{\hbar^2} \over {L^2}} {{4 \pi^2} \over {2 m}}
[n + f]^2
\end{equation}

\noindent where $n = 0, \pm 1, + 2\dots $

Since these levels are periodic in flux, with period $\Phi_0$, the early
presumption was that current, magnetization, and other properties necessarily
be periodic functions with period $\Phi_0$. While this is true, one
cannot exclude the possibility of periodicities with smaller period due to
other reasons, for example, due to electron-electron interaction.

For more than one electron there may be a redistribution of electrons in
energy levels, to gain the lowest ground state energy, when there are
degenerate level crossings. For example one can place two electrons in the
state $n=0$. If there is a total of three electrons, the third electron,
for $f\sim 0$, can be placed either in the $n=1$ state or the $n=-1$ state,
depending on whether $f$ is infinitesimally negative or positive. These
rearrangements, when permited by Fermi statistics, lead to discontinuous
jumps in the current, with changes of sign. Repeating the analysis for the
cases $N = 4 n , 4 n + 1$, $4 n + 2$, $4 n + 3$ leads to the results
given in reference 14.

For odd number of electrons it is seen that there is an approximate half
quantum flux periodicity, which however improves as the number of electrons
increases. For even number of electrons there is full flux periodicity.
These two cases taken alone reproduce the original results obtained
for spinless electrons.

Total current from all occupied levels at $T=0$ is given by $I=\sum\limits_n
I_n$ where the sum is over occupied levels on spin states.
Note, that the main contribution to the sum is always from an electron,
located on the uppermost level. If
$T\not= 0$ it is first necessary to calculate the chemical potential, at
each flux value, as described in the introduction. Thereafter the
$T\not= 0$ result can be found from
$$I= \sum_n I_{n\sigma} f (E_{n \alpha})$$

\noindent where $I_{n\sigma}$ are the different current contributions and
$f(E_{n\sigma})$ is the Fermi function which is determined after the
chemical potential is calculated. The sum over states is taken until there
is a negible remaining contribution. For the exited states there is
greater freedom of  redistributing electrons among levels. Since higher
levels contribute to opposite signs of current one can ask whether there
are possible changes of sign. The periodicity of the Chemical potential,
which is that of one flux quanta, also influences results.

Disorder is introduced in the tight binding formulation in the form of the
usual Anderson type model, for the single site energies along the
rings, with hopping parameter $t$. Thus there is disorder  taken from an
uniform distribution of energies in $[-{W\over{2}}\ ,\ {W\over {2}}]$. Flux
is introduced in the usual way by multiplying the hopping parameters by a
phase factor $e^{i\phi}$ for anticlockwise motion or by $e^{-i\phi}$ for
clockwise motion along the ring, where $\phi = {{2 \pi} \over N}f$.

Current can be either calculated from the slope of the eigenvalues, as a
function of flux, in the usual way; or from (see ref. 17,18).
\begin {eqnarray}
I_m = \left (\frac {2ae}{ {\hbar}}\right ) Im \left (C^{*(m)}_{n,\sigma}
                 C^{(m)}_{n-1,\sigma} H_{n,n-1}\right )
  \nonumber
              = {\frac {2ae} { {\hbar}} Im (C_{n,\sigma}^* C_{n-1,\sigma}
                 t e^{-i\phi})}
\end{eqnarray}
\noindent where $a = {L \over N}$ is the lattice constant.

This expression is actually independent of site number $n$. The
$C^{(m)}_{n,\sigma}$
are the normalized eigenvectors of the diagonalized Hermitian matrix
representing
the Hamiltonian for the $m^{th}$ state and $n^{th}$ site and spin $\sigma$.

Both methods are equivalent, but the slope method is more commonly used, since
eigenvalues are determined more accurately. For zero temperature calculation,
at
each separate flux the energy levels must be filled with the fixed number
of elecrons, starting from the bottom, according to Fermi statistics, to get
the total current from individual currents of the separate levels. This
procedure garantees that the ground state is always achieved, with
readjustments in the level filling at degeneracies described earlier, thus
avoiding a common mistake.

It has been shown that the half quantum flux period arises
by averaging over many rings in situations in which there is
only single quantum flux periodicity for single rings.
Montambaux et al.\cite{19} \cite{20} average over
an ensamble of rings having different
number of electrons to get the
half quantum flux period. We also find  that
this is a good way half-flux quantum periodicity. Averaging over
the four separate cases $N = 4 n ,\ 4 n + 1 ,\ 4 n + 2,\ 4 n + 3$ is one
way to accomplish this periodicity, even without disorder. The most realistic
way
to average is  however
to keep the number of electrons relatively constant.
For strong disorder the sign of the current is sample
dependent and cannot be predicted for a single ring, but
will have the period of just one quantum flux. The emergence
of the half quantum flux on averaging is due to the random
phase.
$(N>>\Delta N$, where $\Delta N$ is the width of the distribution).
\bigskip

\section{ RESULTS AND DISCUSSION}

For both disorder and finite temperature one finds that regions near
$f = 0$ and $f = {1/{2}}$, are with degenerate level crossing.
The latter are very
sensitive to various physical effects.
In these regions, instead of sudden jumps,
the current will become a continuous function of flux, passing through
zero, as soon as there is finite disorder or temperature.

We have noted earlier that for a small odd number of electrons the current
is only a quasi half-periodic function of flux, with a slightly asymetric
shape.  For 5 electrons the current vanishes at $f = .2$ instead of
$f = .25$, for example.

When the ring is disordered, the current changes over to a full flux
periodicity $\Phi_0$ us shown in Figure 1. For a small number of odd
electrons the phase of the $\Phi_0$ period has a definite sign, to accomodate
the direction of the largest current that one obtains in the ordered case.
For a larger odd number of electrons one cannot a-priori predict the phase
of the cycle. Note that the quasi half-flux period has dissapeared by
adding disorder, for individual rings. This behaviour is similar to
that found in simulations on rings with or without leads\cite{1-12}  with
independent electrons.

Let us note that on a single ring, at fixed value of flux, the direction of
current changes upon adding one electron.  The
uppermost level plays the main role in determining the direction and
the sign of the current. Therefore
one expects that processes which move an electron from one level to another,
can change the sign of the current. Such situations may arise with finite
temperature, or disorder. For finite temperature there will be an activation
energy $\Delta\ E\sim T$ for the process. The upper level  has less occupation,
but if the difference of current for a single electron transfered is large
enough, there will be changes of sign. Similarly it is clear that disorder may
play a similar roll, where one has fluctuations in the internal potential
instead of temperature fluctuations.

The finite temperature results are shown in Figures 2a and 2b, both as a
function of flux $f$ (for a fixed temperature) and as a function of
$\beta t$ for a particular value of $f$. Results show how the current changes,
but the shape is preserved, and how there is a change of sign as
temperature is increased. This change of sign occurs when the current
is already rather small.  Also we neglect inelastic events in the calculation.
However this change of sign may perhaps be subject to experimental observation;
detailed temperature studies for isolated individual rings is still lacking.
In the limit of low temperature figure 2b shows a fine step structure.
We believe that this structure reflects the discreetness of the energy
levels for these small systems. The structure dissapears at higher
temperatures.

Let us now see how we can reconcile some of the results with experiment.
Recent experiments\cite{21} have ascribed only the full quantum period in
the case of individual rings.

The easiest way to understand this is either

\begin{itemize}

\item{(a)} That only individual independent single  electrons (i.e. the
one electron case) contributes to the result; or

\item{(b)} That the individual rings are sufficiently disordered to
show only the full quantum periodicity.
\end{itemize}
Averaging over the four cases indicated, inmediately gives the $\Phi_0/2$
period. This interpretation is then natural for the experimental results
concerning many rings\cite{22}. It is interesting to note also that in
the ordered limit the cases with even number of electrons carry a greater
current than cases with odd number of electrons. Where the original
problem had been to explain $\Phi_0/2$ periodicity, it seems that now
there are too many ways to get it. The problem now seems to be why the
$\Phi_0$ period is seen in individual rings.

Recently it has been argued that correlations, being important, cannot
be ignored.  Note that for the spinless fermions
the correlations do not destroy the parity effect,  as was first shown
by Kusmartsev \cite {23},\cite{24}   with the aid of Bethe ansatz
for the limit of strong interactions. This was generalized
by Legett (Legett conjucture) \cite{25} and proven
by Daniel Loss\cite{26} with the aid of bosonisation method
for arbitrary coupling.

For the case of electrons with spin, the trouble is, that these
correlations easily
destroy that parity effect and  create
the half-quantum period \cite {27},\cite{28}  or more generally the
fractional $\Phi_0/N$ periodicity
for $N$ electrons
\cite{29}
and confirmed in other studies \cite{30}.  Recently, we have found that this
fractional  Aharonov-Bohm effect may exist for any coupling in
dilute systems.\cite{31}
Why these effects are not yet seen
in individual rings remains an open problem. Correlations are just one
extra factor to be considered, but so is the problem of multiple channels
expected for rings of finite cross-section.

However, for a  ring of noninteracting electrons with spin,
in the presence of just weak disorder, parity effects may
exist for a small number of electrons and
has the characteristics which we describe. It is relevant to the
phenomenon of the directional changes of the
persitent current with the temperature.
For strong disorder the direction of current cannot be predicted
and is essentially random and sample dependent for individual
rings. Parity effects
are then seemingly inconsquential in this limit, but the situation
if further discussed in detail below.

\section{ DOUBLE PARITY EFFECT}

The essence of parity effect for spinless interacting fermions,
lies in  Fermi statistics \cite {23}-\cite{26}.
 When the number of spinless fermions on the ring
 changes from odd to even, there is
 a statistical half-flux quantum which shifts the energy-flux
dependence  by exactly half of the fundamental flux quantum. Therefore,
 for small values of the flux and at odd number of spinless
fermions, the ring behaves as diamagnet. When there is an even number
of particles it behaves a paramagnet.  Kusmartsev obtained this result by
exact solution with the aid of Bethe -ansatz, in the model of
interacting spinless fermions
on the ring \cite{23} \cite{24} .  This was also independently qualitatively
discussed by Legett  for general case ( called as Legett conjucture) \cite {25}
 and   was proven
by Daniel Loss \cite {26}  with the aid of bosonisation
method in the framework  of
the same model \cite {23} \cite{24}  but for arbitrary coupling.
Thus, the difference in paramagnetic and diamagnetic responses is related to
the
statistical flux on the ring.

The parity effect also occurs because the energy position of uppermost
filled level has the most importance. With a new incoming electron, a new level
is occupied and the  persistent
current  changes direction. However at finite temperature and/or
 with finite  disorder there occurs an intermixing
between levels and it seems that
the parity effect disappears. In fact this is
not quite correct, the parity effect does not disappear.
 The temperature as well as a disorder makes  the single flux periodical
energy-flux dependence  smoother.  That is, with the
temperature or with disorder
the persistent current-flux dependence
takes a form  similar to a $
{\bf \sim sin(2 \pi f) }$
 curve.
The parity effect shows up as a shift of this curve by $\pi$
with the change of the number spinless
fermions. In other words,
 the half quantum flux shift of
the energy-flux dependence with the change from even to an odd or from an odd
to
an even number of spinless fermions does not disappear.

 To take a specific example, with the temperature and at odd number
of fermions on the ring the persitent current first
decreases  from zero and then increases
to zero when the flux changes from zero to 1/2 of the
fundamental flux quantum. For the next
half flux quantum $ 1/2<f<1$
the current  first increases from zero then decreases to zero. On the other
hand for even number of fermions on the ring
for the first half flux quantum $  0<f<1/2$
current  first increases from zero and then
decreases to zero and for the next
half flux quantum $ 1/2<f<1$ the persitent
current first decreases  from zero and then increases
to zero . This is an example of the appearence of
the parity effect for the ring with
 spinless fermions at nonzero temperature or with the presence of disorder.

Let us now take the case of electrons with spin.
As  discussed above, due to

Fermi statistics and due to the intersection of
four levels at zero flux ( for
spinless fermions there was intersection of two levels)
there occurs four
cases, associated with number of electrons $N = 4 n-1  , 4 n $, $4 n +1$, $4 n
+ 2$,
where the behavior of persistent current is distinct. The
cases with $N=4n-1$ and $N=4n+1$  resemble the case of spinless fermions.
 For these cases the ground state energy-flux dependence is shifted by
half a quantum flux.  The same situation occurs for the other two cases
of $N=4n$ and $4n+2$ number of electrons on the ring.

That is, for the single ring with free electrons we have two  parity effects
which occur when the number of electrons changes by 2.
When the number of electrons
changes from $N=4n-1$ to $N=4n$ the shape of the
ground state energy-flux dependence changes
gradually, becoming a quasi-half flux periodical function.
The dependence of the persistent current
on the flux $f$ ( $0<f<1/2$)  in this case consists
of two nonequivalent half-periods. One is much smaller than the second one.
This nonequivalence  for the small amplitude  half-period which occurs
at small flux,
arises  in the case when three of four
intersecting   levels are filled by electrons, that is when
$N=4n+1$. It is important to not that in all
cases here, except when all intersecting levels are
filled  the paramagnetic response occurs.
For the cases $N=4n+2$ the response will have
a diamagnetic character.
The same  effect occurs when the number of electrons change
from $N=4n-1$ to $N=4n$, where the behavior
similar to discussed above but with the
shift of a half flux quantum.
 The reason for such behavior that  the addition a of new electron
creates the statistical half-flux quantum.

To conclude with the case of electrons with spin, the parity effect also
exists,
but takes a new
form: instead of two types of ground state energy-flux
dependence for  spinless
fermions, which are related to an even and odd number of particles,
 there are four different types
 of the ground state energy-flux dependence.

However with finite disorder
or finite temperature this parity effect changes in character.
With finite temperature the energy region
of  intersecting levels becomes equally populated .
Let us consider each of four cases independently.
When there is only one or two
electrons on the four intersecting levels, then,  with
a slight increase of the temperature
all four levels will be populated and
the current  will be changed only in the flux region
where these levels are near each other.
The absolute value of that current strongly
decreases resulting in a smooth behavior.
With the increase of flux $f$, the distance between
levels $\Delta(f)$ increases .
One can consider that the temperature $T$ is switched off, if one
has $T< \Delta(f)$ and will be swiched on only
in the flux region of next four intersecting levels, where $\Delta(f)<T$.

Now let us consider the case
when there are three electrons on  four intersecting levels. In that case
a small amplitude half period occurs in the flux dependence of
persistant current( in the flux region $0<f<.2$).
At finite temperature all these four levels
will be equally populated, with  vanishing resulting current.
As a result the original single flux periodicity will appear.
Thus, the reason
why the single flux periodicity occurs with finite temperature or  disorder is
the intersection of levels and parity effect, caused by Fermi statistics.

Let us discuss what kind of parity effect we have at finite temperature.
As we have discussed above the cases
with $N=4n-1$ and $N=4n$ electrons behave similarly
to each other. In both of these cases the response is paramagnetic. The other
two self-similar cases  associated with  $N=4n+1$
and $4n+2$ electrons on the ring.
But the response of the latter two cases has a
diamagnetic character. The dependence
of the persitent current on the flux for
the first  group may be obtained from the second one
with a shift of a half-flux quantum. Thus
the parity effect might be seen by comparing
these two different types of the flux-current dependence.

However, the Hubbard like
interaction may destroy the parity effect with the creation of fractional
Aharonov-Bohm effect \cite{29}-\cite{31}.  As described
by Haldane,  fractional statistics
or statistical flux may appear with the interaction \cite {32}.
In this case the interaction
creates the needed half-flux quantum to
compensate the statistical flux which occurs
due to the parity
effect.

\section{ RECENT EXPERIMENT}

Recently , there has been an experiment
on a semiconductor single loop in the $GaAs/GaAlAs$ system \cite {33}
for which single flux periodicity has been detected.
After an analysis of the experimental data it was strongly concluded
\cite {21} that
in present theories the disorder is not correctly taken into account.
Since the loop has only a few electron channels
( it was estimated as equal to 4), our theory
may be applicable:

1) The single flux periodicity is due to disorder, as described in the
text

2) The persistent currents are sample specific. As we have
shown, the persistent current may change even in sign, with the change
of the  level of disorder.

To complete a confirmation of our predictions ( to observe the
temperature changes of the sign of persistent current)
 one needs to measure
the persistent current at distinct values of temperatures and of flux.
That such measurements are needed, was also concluded in Ref.
\cite {33}.
 It is worth noting that the findings are also  applicable to
the description of the frequency changes of phonons on the ring
in magnetic field (see, Ref.\cite {13}) with temperature.

\vskip 1truecm
\parindent 20truept

\leftline{ACKNOWLEDGEMENTS}

F.V.K. and J.F.W. wish to thank INPE and CNPQ for financial support during
their stay at INPE, Sao Jose dos Campos, S.P., Brasil. FK thanks
D. Mailly for the discussion of experimetal results \cite {33} and
Ministry of Education, Science and Culture, Japan for partial support.
J.F.W. also thanks foundations ANDES/ANTORCHAS/VITAE for grant support
In the early stages the current calculating program was written along
with Erasmo A. de Andrada e Silva, who is currently in Pisa, Italy.
\vskip 1truecm

\leftline{REFERENCES}

(*) permanent address: L.D.Landau Institite, Moscow, Russia;

\vskip 1truecm
\leftline{Figure Captions}

{Figure 1}{:}{Persistant current for an ordered (Fig. 1a) and
disordered (Fig 1b) ring with 5 electrons and 40 sites. In the
ordered case current vanishes at $f = .20$ rather than $f=.25$. For
the disorder case a parameter $W = .4 t$ was used, with the Anderson
model. Note change in sign of current for some values of flux.
With disorder the quasi half quantum flux period has dissapeared.

{Figure 2a}{:}{Current at a fixed finite temperature ($\beta t = 30$) as
a function of flux, for an isolated ring of 41 electrons and 40 sites
current is zero at degenerate level crossing points  $f=0$ and
$f={1\over {2}}$. Additionally the periodicity of half a quantum flux
for the zero temperature ordered chain is maintained.
\bigskip

{Figure 2b}{:}{Current as a function of temperature ($\beta t$ variable)
for a fixed value of flux ($f = .23$).  A change of sign is seen at high
temperature, region in which the current is small. The change in sign
is due to exitation to higher levels, which carry current of opposite
sign. Additional structure due to discretness of levels is also seen,
which is smoothed out at higher temperature.

\end{document}